\documentclass[]{aastex701}

\usepackage{xcolor}

\begin{document}

\title{Multi-line Wave Signatures in a Sunspot from Near-UV {\sc Sunrise iii} /SUSI Observations}

\author[orcid=0000-0002-7711-5397]{Shahin Jafarzadeh}
\affiliation{Astrophysics Research Centre, School of Mathematics and Physics, Queen’s University Belfast, Belfast, BT7 1NN, UK}
\affiliation{Max Planck Institute for Solar System Research, Justus-von-Liebig-Weg 3, 37077 G\"{o}ttingen, Germany}
\email[show]{shahin.jafarzadeh@qub.ac.uk}  

\author[orcid=0000-0002-9155-8039]{David B. Jess} 
\affiliation{Astrophysics Research Centre, School of Mathematics and Physics, Queen’s University Belfast, Belfast, BT7 1NN, UK}
\affiliation{Department of Physics and Astronomy, California State University Northridge, Northridge, CA 91330, USA}
\email[]{}  

\author[orcid=0000-0002-5365-7546]{Marco Stangalini} 
\affiliation{ASI Italian Space Agency, Via del Politecnico snc, I-00133 Rome, Italy}
\email[]{}  

\author[orcid=0000-0001-5678-9002]{Richard~J.~Morton} 
\affiliation{School of Engineering, Physics and Mathematics, Northumbria University, Newcastle upon Tyne, NE1 8ST, UK}
\email[]{}  

\author[orcid=0000-0003-1732-6632]{Tobias Felipe}
\affiliation{Instituto de Astrofísica de Canarias, E-38205 C/ Vía Láctea, s/n, La Laguna, Tenerife, Spain}
\affiliation{Departamento de Astrofísica, Universidad de La Laguna 38205, La Laguna, Tenerife, Spain}
\email[]{} 

\author[orcid=0009-0007-2465-1931]{Michele Berretti}
\affiliation{University of Trento, Via Calepina 14, 38122 Trento, Italy}
\affiliation{University of Rome Tor Vergata, Department of Physics, Via della Ricerca Scientifica 3, 00133 Rome, Italy}
\email[]{} 

\author[orcid=0000-0002-3418-8449,sname='Solanki']{Sami~K.~Solanki} \affiliation{Max Planck Institute for Solar System Research, Justus-von-Liebig-Weg 3, 37077 G\"{o}ttingen, Germany}\email{solanki@mps.mpg.de}	

\author[orcid=0000-0003-3490-6532,sname='Smitha']{H.~N.~Smitha} \affiliation{Max Planck Institute for Solar System Research, Justus-von-Liebig-Weg 3, 37077 G\"{o}ttingen, Germany}\email{narayanamurthy@mps.mpg.de}	

\author[orcid=0000-0003-1459-7074,sname='Lagg']{Andreas~Lagg} \affiliation{Max Planck Institute for Solar System Research, Justus-von-Liebig-Weg 3, 37077 G\"{o}ttingen, Germany}\email{lagg@mps.mpg.de}	

\author[orcid=0000-0002-9972-9840,sname='Gandorfer']{Achim~Gandorfer} \affiliation{Max Planck Institute for Solar System Research, Justus-von-Liebig-Weg 3, 37077 G\"{o}ttingen, Germany}\email{gandorfer@mps.mpg.de}

\author[orcid=0009-0009-4425-599X,sname='Feller']{Alex~Feller} \affiliation{Max Planck Institute for Solar System Research, Justus-von-Liebig-Weg 3, 37077 G\"{o}ttingen, Germany}\email{feller@mps.mpg.de}	

\author[orcid=0000-0003-1409-1145,sname='Iglesias']{Francisco~A.~Iglesias} \affiliation{Max Planck Institute for Solar System Research, Justus-von-Liebig-Weg 3, 37077 G\"{o}ttingen, Germany}\affiliation{Grupo de Estudios en Heliofísica de Mendoza, CONICET, Universidad de Mendoza, Boulogne sur Mer 683, 5500 Mendoza, Argentina}\email{iglesias@mps.mpg.de}	

\author[orcid=0000-0001-6317-4380,sname='Riethmüller']{Tino~L.~Riethmüller} \affiliation{Max Planck Institute for Solar System Research, Justus-von-Liebig-Weg 3, 37077 G\"{o}ttingen, Germany}\email{riethmueller@mps.mpg.de}	

\author[sname='Grauf']{Bianca~Grauf} \affiliation{Max Planck Institute for Solar System Research, Justus-von-Liebig-Weg 3, 37077 G\"{o}ttingen, Germany}\email{grauf@mps.mpg.de}	

\author[orcid=0000-0001-6029-7529,sname='Hoelken']{Johannes~Hoelken} \affiliation{Max Planck Institute for Solar System Research, Justus-von-Liebig-Weg 3, 37077 G\"{o}ttingen, Germany}\email{hoelken@mps.mpg.de}
\author[orcid=0000-0002-5054-8782,sname='Katsukawa']{Yukio~Katsukawa} \affiliation{National Astronomical Observatory of Japan, 2-21-1 Osawa, Mitaka, Tokyo 181-8588, Japan}\affiliation{Department of Astronomy, The University of Tokyo, 7-3-1, Hongo, Bunkyo-ku, Tokyo 113-0033, Japan}\affiliation{Department of Astronomical Science, The Graduate University for Advanced Studies (SOKENDAI), 2-21-1 Osawa, Mitaka, Tokyo 1818588, Japan}\email{yukio.katsukawa@nao.ac.jp}	
\author[orcid=0000-0002-0787-8954,sname='Bernasconi']{Pietro~Bernasconi} \affiliation{Johns Hopkins University Applied Physics Laboratory, 11100 Johns Hopkins Road, Laurel, Maryland, USA}\email{pietro.bernasconi@jhuapl.edu}	
\author[sname='Berkefeld']{Thomas~Berkefeld} \affiliation{Institut für Sonnenphysik (KIS), Georges-Köhler-Allee 401a, 79110 Freiburg, Germany}\email{thomas.berkefeld@leibniz-kis.de}	
		
\author[orcid=0000-0001-9228-3412,sname='Álvarez-Herrero']{Alberto~Álvarez-Herrero} \affiliation{Instituto Nacional de T\'ecnica Aeroespacial (INTA), Ctra. de Ajalvir, km. 4, E-28850 Torrejón de Ardoz, Spain}\affiliation{Spanish Space Solar Physics Consortium}\email{alvareza@inta.es}	
\author[orcid=0000-0001-5616-2808,sname='Kubo']{Masahito~Kubo} \affiliation{National Astronomical Observatory of Japan, 2-21-1 Osawa, Mitaka, Tokyo 181-8588, Japan}\email{masahito.kubo@nao.ac.jp}	
\author[orcid=0000-0001-8829-1938,sname='Orozco~Suárez']{David~Orozco~Suárez} \affiliation{Instituto de Astrofísica de Andalucía, CSIC, Glorieta de la Astronomía s/n, 18008 Granada, Spain}\affiliation{Spanish Space Solar Physics Consortium}\email{orozco@iaa.es}	
\author[sname='Carpenter']{Michael~Carpenter} \affiliation{Johns Hopkins University Applied Physics Laboratory, 11100 Johns Hopkins Road, Laurel, Maryland, USA}\email{michael.carpenter@jhuapl.edu}	
\author[sname='Bell']{Alexander~Bell} \affiliation{Institut für Sonnenphysik (KIS), Georges-Köhler-Allee 401a, 79110 Freiburg, Germany}\email{albe@leibniz-kis.de}	
\author[orcid=0000-0001-7764-6895,sname='Martínez~Pillet']{Valentín~Martínez~Pillet} \affiliation{Instituto de Astrofísica de Canarias, Vía Láctea, s/n, E-38205 La Laguna, Spain}\affiliation{Spanish Space Solar Physics Consortium}\email{vmpillet@iac.es}
		
\author[orcid=0000-0002-7318-3536,sname='Bailén']{Francisco~Javier~Bailén} \affiliation{Instituto de Astrofísica de Andalucía, CSIC, Glorieta de la Astronomía s/n, 18008 Granada, Spain}\affiliation{Spanish Space Solar Physics Consortium}\email{fbailen@iaa.es}	
\author[orcid=0000-0002-2055-441X,sname='Blanco~Rodríguez']{Julian~Blanco~Rodríguez} \affiliation{Universitat de Valencia Catedrático José Beltrán 2, E-46980 Paterna-Valencia, Spain}\affiliation{Spanish Space Solar Physics Consortium}\email{julian.blanco@uv.es}	
\author[orcid=0000-0003-4319-2009,sname='Castellanos~Durán']{Juan~Sebastián~Castellanos~Durán} \affiliation{Max Planck Institute for Solar System Research, Justus-von-Liebig-Weg 3, 37077 G\"{o}ttingen, Germany}\email{castellanos@mps.mpg.de}	
\author[orcid=0009-0002-6808-5154,sname='Harnes']{Edvarda~Harnes} \affiliation{Max Planck Institute for Solar System Research, Justus-von-Liebig-Weg 3, 37077 G\"{o}ttingen, Germany}\email{harnes@mps.mpg.de}	
\author[orcid=0000-0002-4669-5376,sname='Ishikawa']{Ryohtaroh~T.~Ishikawa} \affiliation{National Institute for Fusion Science, 322-6 Oroshi-cho, Toki City 509-5292, Japan}\email{ishikawa.ryohtaro@nifs.ac.jp}	
\author[orcid=0000-0001-7452-0656,sname='Kawabata']{Yusuke~Kawabata} \affiliation{National Astronomical Observatory of Japan, 2-21-1 Osawa, Mitaka, Tokyo 181-8588, Japan}\email{kawabata.yusuke@nao.ac.jp}	
\author[orcid=0000-0002-1043-9944,sname='Matsumoto']{Takuma~Matsumoto} \affiliation{Centre for Integrated Data Science, Institute for Space-Earth Environmental Research, Nagoya University, Furocho, Chikusa-ku, Nagoya, Aichi 464-8601, Japan}\email{takuma.matsumoto@gmail.com}	
\author[orcid=0000-0002-7044-6281,sname='Oba']{Takayoshi~Oba} \affiliation{Advanced Research Center for Space Science and Technology, Institute of Science and Engineering, Kanazawa University, Kakuma-machi, Kanazawa, Ishikawa 920-1192, Japan}\affiliation{Max Planck Institute for Solar System Research, Justus-von-Liebig-Weg 3, 37077 G\"{o}ttingen, Germany}\email{oba@mps.mpg.de}	
\author[orcid=0000-0003-0175-6232,sname='Siu-Tapia']{Azaymi~L.~Siu-Tapia} \affiliation{Instituto de Astrofísica de Andalucía, CSIC, Glorieta de la Astronomía s/n, 18008 Granada, Spain}\affiliation{Spanish Space Solar Physics Consortium}\email{siu@iaa.es}	
\author[orcid=0000-0003-1483-4535,sname='Strecker']{Hanna~Strecker} \affiliation{Instituto de Astrofísica de Andalucía, CSIC, Glorieta de la Astronomía s/n, 18008 Granada, Spain}\affiliation{Spanish Space Solar Physics Consortium}\email{streckerh@iaa.es}	
\author[orcid=0000-0003-1971-5551,sname='Vukadinović']{Dušan~Vukadinović} \affiliation{Institut für Physik, Universität Graz, Universitätsplatz 5, 8010 Graz, Austria}\affiliation{Max Planck Institute for Solar System Research, Justus-von-Liebig-Weg 3, 37077 G\"{o}ttingen, Germany}\email{vukadinovic@mps.mpg.de}	

\begin{abstract}
Magnetohydrodynamic waves redistribute energy in magnetic structures of the lower solar atmosphere, yet constraints on how wave power and dominant frequencies are organised above sunspots remain limited because most studies use only a few well-separated diagnostics. Here we present multi-line wave signatures in a sunspot from near-UV spectroscopy with the \textsc{Sunrise-iii} UV Spectropolarimeter and Imager (SUSI). We analyse a two-hour time series of repeated raster scans of a sunspot near disc centre in the 327--329\,nm spectral window ($>100$ lines). From these, we select 44 lines that radiative-transfer calculations suggest sample effective formation heights within the umbral core from deep photosphere toward the low chromosphere. For each line, we extract line-core intensity and line-of-sight velocity time series using a dedicated multi-line fitting routine and compute Morlet-wavelet power spectra. The refined global wavelet spectra show that most lines (in both intensity and velocity) are genuinely multi-frequency, with a dominant peak and substantial statistically significant power up to 12\,mHz. Unsupervised clustering of the normalised spectra groups lines into families with similar spectral shapes and reveals a progression of dominant frequencies from $\sim$2 to $\sim$10\,mHz across the ensemble, for both intensity and velocity (not necessarily in the same lines). This behaviour is not reproduced by a simple formation-height ranking, suggesting that uncertainties in the formation-height estimates and line-dependent diagnostic response together shape the ordering. These \textsc{Sunrise-iii}/SUSI observations open a new regime for near-UV multi-line wave studies and provide the first systematic characterisation of frequency-structured sunspot wave behaviour in this spectral region.
\end{abstract}

\section{Introduction} 
\label{sec:introduction}

Energy transport and heating in the solar atmosphere are believed to be strongly influenced by magnetohydrodynamic (MHD) waves, especially in magnetised regions such as network, plage, pores, and sunspots \citep[e.g.,][]{2005LRSP....2....3N, 2015SSRv..190..103J, 2023LRSP...20....1J, 2021RSPTA.37900172G}. In the lower atmosphere, waves are guided by structured magnetic fields, and partially converted between different MHD modes when they encounter layers where the Alfvén and sound speeds are comparable \citep{1994ApJ...437..505C, 2023ApJ...946..108C}. Observationally, this gives rise to a rich spectrum of intensity and line-of-sight (LOS) velocity oscillations whose properties vary with height, position, and magnetic topology \citep[e.g.,][]{2022ApJ...938..143G}. Disentangling the roles of atmospheric stratification, magnetic geometry, and mode conversion in shaping the observed oscillatory spectra requires spatially and spectrally well-resolved, multi-height observations spanning the photosphere and chromosphere.

Sunspots provide a natural laboratory for such studies. Their strong, coherent magnetic fields form large-scale waveguides, while the sharp contrasts between umbra, penumbra, and surrounding quiet Sun give rise to a variety of wave phenomena throughout the atmosphere, including umbral flashes, running penumbral waves, and global eigenmodes \citep{2000SoPh..192..373B, 2006RSPTA.364..313B, 2015LRSP...12....6K, 2018NatPh..14..480G}. Previous works have used photospheric and chromospheric diagnostics, such as Fe~{\sc i}, Ca~{\sc ii}, H$\alpha$, and He~{\sc i}, and molecular CO lines, to explore wave propagation along sunspot magnetic fields, to identify dominant frequencies and eigenmodes, and to infer the presence of resonance cavities \citep[e.g.,][]{1996A&A...315..303S, 2006ApJ...640.1153C, 2007ApJ...671.1005B, 2010ApJ...722..131F, 2015ApJ...812L..15K, 2020NatAs...4..220J, 2021RSPTA.37900216S}. However, most of these studies rely on a relatively small set of spectral lines, often separated by several hundred kilometres in formation depth/height, which limits the ability to track how wave properties can reorganise over relatively short height ranges in response to strong stratification and changing magnetic geometry (e.g., rapid variations in density/temperature scale heights, field inclination, and the local cut-off and transmission conditions).

From a theoretical perspective, the dominant frequencies and wave power are expected to change with height for several reasons. The effective acoustic cut-off frequency depends on temperature and field inclination, so slow magnetoacoustic waves guided along increasingly inclined fields can leak low-frequency wave power into higher layers, while more vertical fields tend to favour higher-frequency propagation \citep[e.g.,][]{2018A&A...617A..39F, 2019ApJ...884L...8J}. Sharp gradients in atmospheric parameters can form resonance cavities between, for example, the temperature minimum and the transition region, selecting particular frequencies and enhancing their power at specific heights \citep[e.g.,][]{2019A&A...627A.169F, 2020NatAs...4..220J}. Mode conversion near equipartition layers can further redistribute energy between various MHD modes, modifying the observed power spectra with height \citep{2006MNRAS.372..551S, 2007AN....328..286C}.

An additional complication is that the observed oscillatory signal is filtered not only by the atmospheric structure, but also by the diagnostic properties of the spectral lines used. Different lines respond to different combinations of perturbations -- temperature, density, velocity, and magnetic field -- through their contribution functions and source-function behaviour \citep{1992ApJ...398..375R, 2023A&A...669A.144S}. Lines with strong temperature sensitivity may preferentially highlight thermal oscillations, while lines affected by scattering or non-LTE effects can add complexity due to the MHD signatures being weighted (or filtered) by the radiative properties of the spectral line, even if multiple lines have similar nominal formation heights \citep{2008A&A...480..515C, 2012ApJ...749..136L}. 
Exploiting a large set of diagnostic spectral lines with diverse sensitivities is thus essential for building a more complete picture of wave behaviour.

Recent high-resolution studies in the lower solar photosphere have revealed that multiple, narrow-band frequency components can coexist within the same magnetic structures, and that these are consistent with distinct MHD eigenmodes supported by the local topology \citep{2022NatCo..13..479S, 2023A&A...674A.109C, 2024A&A...688A...2J}. Such results highlight the need for densely sampled, multi-height observations capable of tracking how these multi-mode wave fields evolve with height, and how their relative power is redistributed between photospheric and chromospheric layers. Extending this type of analysis into previously unexplored spectral regimes offers the prospect of identifying additional wave modes and constraining their coupling across heights.

The near-ultraviolet (near-UV) part of the spectrum offers an attractive but historically under-investigated window for such wave studies. In this spectral region, the solar spectrum is crowded with photospheric and low-chromospheric lines from a range of neutral and ionised species \citep{1973apds.book.....D}. Many of these lines have not previously been exploited, particularly for detailed oscillation diagnostics, largely because of the observational challenges associated with ground-based near-UV spectroscopy. At the same time, the dense distribution of lines in this region provides the potential for a finely sampled height stratification of diagnostics, provided that representative formation heights (or depths relative to a reference atmosphere) can be estimated and that the line profiles can be measured with sufficient spectral and temporal resolution.

Following the two successful earlier flights of the {\sc Sunrise} balloon-borne solar observatory in 2009 and 2013 \citep{2010ApJ...723L.127S, 2011SoPh..268....1B, 2017ApJS..229....2S}, the third flight in July~2024 opened this regime for the first time with high spatial, spectral, and temporal resolution. \textsc{Sunrise iii}  carries a 1-m aperture telescope and three polarimetric instruments designed to probe the lower solar atmosphere in seeing-free conditions \citep{2025SoPh..300...75K, solankietal2026}. The {\sc Sunrise} UV Spectropolarimeter and Imager (SUSI) is a slit-based spectropolarimeter operating in the 309–417\,nm range, optimised for simultaneous measurements of many near-UV lines with full Stokes polarimetry and high cadence \citep{2025SoPh..300...65F, 2025SoPh..300...58I}.

Here, focus is placed on a single SUSI dataset: a two-hour time series of a large sunspot near disc centre, observed in the 327--329\,nm spectral window. This window contains more than a hundred identifiable lines with model-estimated formation heights (referenced to quiet-Sun $\tau_{500\,\mathrm{nm}}=1$) spanning from the deep photosphere to the low chromosphere. For a carefully selected subset of 44 relatively clean lines, line-core intensity and LOS velocity are extracted as a function of time and position along the slit, enabling a systematic investigation of how wave signatures vary across a dense set of diagnostics and their inferred formation heights inside the sunspot. In this Letter, only the SUSI Stokes~$I$ spectra are used; the polarimetric information will be exploited in future work once full Stokes products become available.

The main aim of this Letter is to provide a first observational characterisation of multi-line wave signatures in a sunspot using the near-UV spectral lines. The analysis concentrates on power spectra of intensity and LOS velocity in the umbra's central region, where the magnetic field is expected to be relatively vertical and height comparisons are therefore most straightforward.

\section{Observations} 
\label{sec:observations}

The data analysed in this study were obtained during the main science phase of the \textsc{Sunrise iii}  flight on 14 July 2024 between 21:40--23:39~UTC. SUSI was operated in a small-raster mode on the 327--329\,nm spectral window, targeting a large sunspot (NOAA\,AR~13744) located close to disc centre ($\mu\approx0.96$). The SUSI spatial sampling is \(0.03^{\prime\prime}\,\mathrm{px}^{-1}\), with spectral sampling in this wavelength range of $\sim8.3\,\mathrm{m\AA\,px^{-1}}$ \citep{2025SoPh..300...65F}.
The SUSI slit was positioned such that the raster crossed the centre of the umbra, extended across the penumbra, and reached into the surrounding plage and a quieter region, while also crossing a nearby pore. Each raster step covered a field of view of approximately \(46\arcsec\) along the slit and $\sim$\(3.5\arcsec\) in the scan direction after processing, producing a narrow but spatially resolved strip across the active region. A full raster cycle in the 327--329\,nm spectral window was completed in about 39\,s, yielding a nearly two-hour time series with 180 scans. A short gap of 156\,s occurs after the first 57\,min of the observations (due to calibrations), but this does not affect the analysis of wave power in the 1–12\,mHz range. The gap is linearly interpolated prior to wave analysis.

In this observing programme, SUSI recorded full Stokes spectra in the selected near-UV window, but the present work only makes use of Stokes~$I$. Standard reduction procedures within the SUSI pipeline were applied to the data, including dark and flat-field corrections, wavelength calibration, and correction for instrumental effects. The slit-jaw images, obtained through the SUSI slit-jaw bandpass filter centred at 325.0\,nm with a full width at half maximum of 0.9\,nm, have been phase-diversity reconstructed, while the spectral scans have not. In a subsequent post-processing step, we constructed time-series raster images from the individual slit positions and applied additional corrections for solar rotation, instrumental image/slit rotation (field derotation), and residual image jitter. The reduced dataset can be represented as a four-dimensional cube with axes corresponding to time, position along the slit, scan position across the raster, and wavelength.

The 327--329\,nm window contains more than one hundred identifiable absorption lines from multiple neutral and ionised species. For the present umbral study, formation heights were estimated at each spatial pixel of a 3D radiative-MHD MURaM sunspot simulation (\citealt{2005A&A...429..335V, 2012ApJ...750...62R}; see also \citealt{2023A&A...669A.144S}) using two complementary radiative-transfer approaches. First, we computed an Eddington–Barbier (EB; \citealt{2021arXiv210302369R}) height-of-formation proxy as the geometric height where $\tau_\lambda=1$ for each wavelength sample. Second, for a smaller set of selected lines we computed contribution functions (CFs) to the emergent intensity. Because CF calculations are computationally expensive, they were performed only for a subset of lines, whereas the EB estimate is available for every wavelength sample across the window. While the two methods generally agree, modest line-dependent differences are expected because EB is an approximation, whereas CF heights incorporate the full opacity and source-function weighting of the line formation. All radiative-transfer calculations in this work were performed assuming local thermodynamic equilibrium (LTE) using the RH code \citep{2001ApJ...557..389U, 2015A&A...574A...3P}, with atomic line parameters taken from the Kurucz line list \citep{1995all..book.....K}.

To obtain representative umbral-core formation heights, we accounted for the SUSI spatial point-spread function (PSF) and averaged over a small umbral-core region. For the CF method, the contribution functions in a $5\times5$ pixel region around the umbral centre were convolved with the SUSI spatial PSF, averaged over the region, and the weighted-mean height of the resulting mean CF was adopted for each line. For the EB method, each wavelength-dependent EB height map was convolved with the same spatial PSF and then averaged over the same $5\times5$ pixel umbral-core region, yielding one representative EB height per wavelength sample. Both height scales are referenced to the quiet-Sun $\tau_{500\,\mathrm{nm}}=1$ level, so the inferred umbral formation heights are mostly negative, reflecting the Wilson depression of the umbral $\tau\approx1$ surface by several hundred kilometres \citep[e.g.,][]{2003A&ARv..11..153S, 2018A&A...619A..42L, 2020A&A...635A.202L}; the diagnostics nevertheless sample a substantial range of layers within the umbral photosphere and toward the low chromosphere. In the wave analysis below we use the CF-based heights, while the EB heights are included for completeness over the full wavelength range. 

In this Letter we adopt the formation-height ranking derived from the 3D sunspot model as a plausible strong-field reference for the umbral conditions studied here, and we use the CF-based heights for the wave analysis. We emphasise, however, that the resulting formation heights remain model dependent and should be regarded mainly as approximate proxies for relative ordering. A more detailed comparison of formation-height estimates obtained from quiet-Sun (FALC), plage (FALP), and the sunspot model atmospheres, and their implications for different magnetic environments (umbra, pore, and surroundings), are presented in \citet{Jafarzadeh2026_paper2}.

\begin{figure*}[!ht]
\centering
\includegraphics[width=1.0\textwidth]{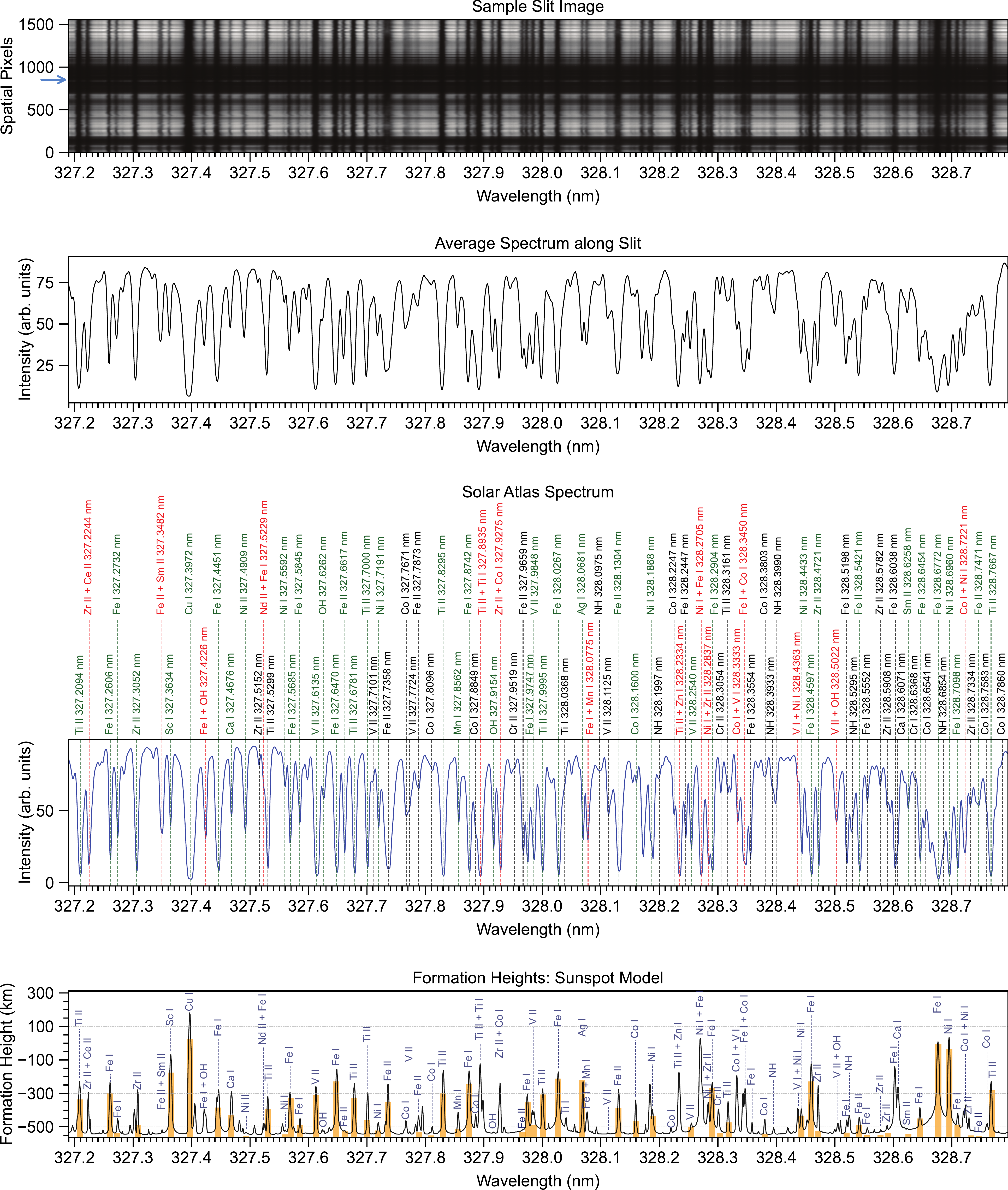}
\caption{Overview of the near-UV SUSI diagnostics used in this study. 
Top: sample SUSI spectrum in the 327--329\,nm window at the beginning of the time series; the blue arrow on the left of the panel marks the slit position (y pixel) used for the umbral-core analysis. 
Second panel: the corresponding average spectrum along the slit, displayed as a function of wavelength to illustrate the overall contrast and line depths. 
Third panel: high-resolution solar atlas spectrum for the same wavelength range, with individual lines identified and labelled; known blended lines are marked in red, the 44 relatively unblended lines selected for the wave analysis are highlighted in green, and the remaining identified lines are shown in black. 
Bottom: Eddington–Barbier (EB) formation heights (solid curve) across the full spectral window, referenced to the quiet-Sun $\tau_{500\,\mathrm{nm}}=1$ level, with contribution-function (CF) heights shown as orange bars for the lines with computed CFs.}
\label{fig:spectra_overview}
\end{figure*}

An overview of the spectral content and the associated diagnostics is shown in Figure~\ref{fig:spectra_overview}. The top panel displays a sample SUSI spectrum recorded at the beginning of the time series, highlighting the dense forest of absorption features in the 327--329\,nm window. The second panel shows the average spectrum along the slit versus wavelength, illustrating the overall contrast and line depths. The third panel presents a corresponding high-spectral-resolution solar atlas spectrum \citep{1966SolarSpectrum}, where individual lines have been identified, labelled, and colour-coded: known blended lines are marked in red, the subset of 44 relatively clean lines selected for the wave analysis is shown in green, and the remaining identified lines are plotted in black. The bottom panel shows the EB formation heights across the full spectral window (solid curve), with the CF-based heights overplotted as orange bars at a selected subset of lines for which CFs were computed.

These observational products, together with the line identification and selection indicated in the atlas panel, form the basis for the multi-line wave analysis presented in the following section, where the extraction of line-core quantities and the construction of the time-series diagnostics are detailed.

\section{Analysis and Results}
\label{sec:analysis}

\begin{figure*}[!t]
\centering
\includegraphics[width=1.0\textwidth]{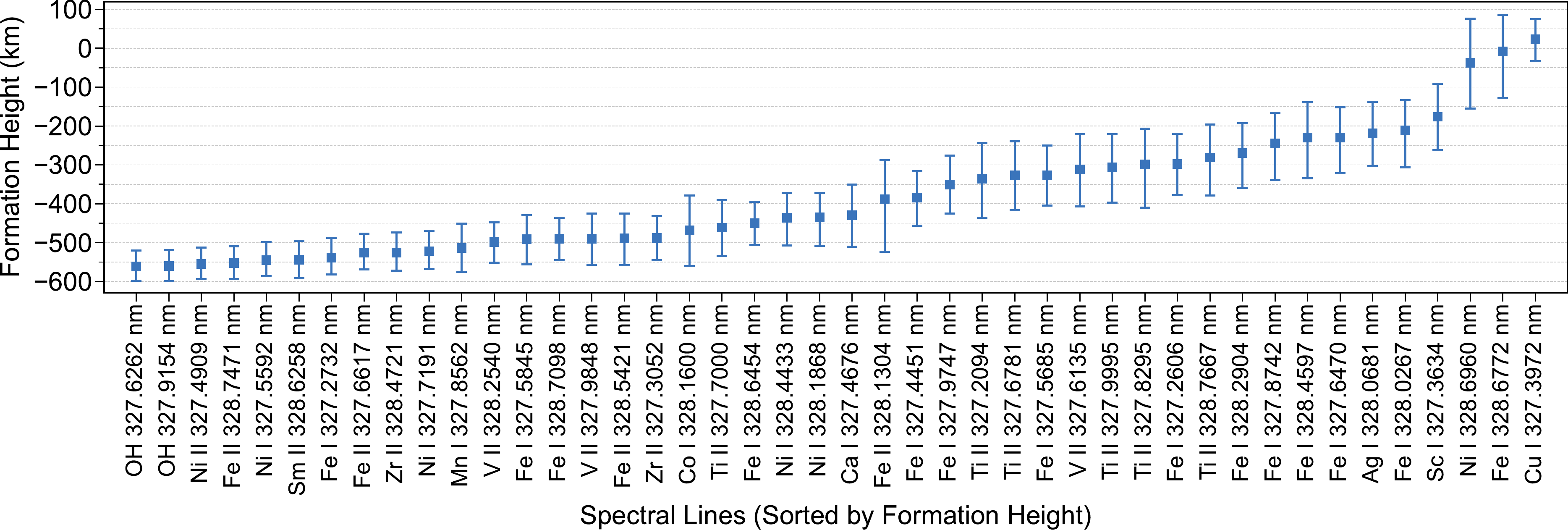}
\caption{Effective formation height ranking for the 44 near-UV spectral lines used in this Letter. The plotted symbols correspond to weighted-mean line-core formation heights derived from contribution functions in a representative 3D radiative-MHD sunspot model, after accounting for the SUSI spatial PSF and averaging over a small umbral-core region. The asymmetric vertical bars show the 16th--84th percentile range of each line's contribution function, illustrating the finite vertical extent of the diagnostic. Heights are measured relative to the quiet-Sun $\tau_{500\,\mathrm{nm}}=1$ level. As a result, most lines appear at negative geometric heights in the umbra, while the relative ranking still spans from deeper photospheric layers toward diagnostics approaching the low chromosphere.}
\label{fig:sunspot_height_ranking}
\end{figure*}

The aim of this work is to quantify how wave signatures vary across the ensemble of near-UV diagnostics sampling a sunspot atmosphere, by exploiting the dense set of spectral lines in the SUSI 327--329\,nm window. From the full set of identified lines, we focus on the subset of 44 relatively unblended lines with clear, well-defined absorption cores (highlighted in green in Figure~\ref{fig:spectra_overview}). These lines were selected based on a combination of line depth, isolation from known blends, and a careful inspection of their umbral profiles, excluding lines that show peculiar or highly dynamic behaviour at the analysis location or strongly skewed cores suggestive of potential blends. Effective line-core formation heights for this subset, obtained from the contribution functions in the 3D sunspot model (including the SUSI spatial PSF and a small umbral-core average; Section~\ref{sec:observations}), are used primarily to provide a first-order stratification and, most importantly, a relative ranking of the diagnostics in terms of sampled height (see Figure~\ref{fig:sunspot_height_ranking}). In addition to the weighted-mean height, we also show the finite vertical extent of each line's contribution function through its 16th--84th percentile range, to illustrate that the diagnostics sample a non-negligible height range that can be asymmetric and overlap substantially with neighbouring lines, rather than a single geometric layer.

In what follows, we employ this ranking to stack and compare wave diagnostics as a function of inferred formation height (noting that heights are referenced to the quiet-Sun $\tau_{500\,\mathrm{nm}}=1$ level and are therefore mostly negative in the umbra), while recognising that the absolute height values are model dependent and subject to uncertainties that will need to be further investigated, and possibly refined, in future modelling work.

\subsection{Multi-line Adaptive Voigt Fitting and Line-core Quantities}
\label{subsec:fitting}

To extract line-core intensity and LOS velocity for each of the selected lines, we employ our dedicated Python-based multi-line fitting routine (\texttt{LineFit}; \citealt{2026FrASS...Jafarzadeh_inprep}) that models each spectral line with a Voigt profile. The code is designed to work on 1D spectra along the slit, but can be applied to all spatial pixels and time steps in a parallelised manner.

For each spectrum, the code first multiplies the intensity profile by \(-1\) (so absorption minima become peaks) and uses a standard peak-finding algorithm to locate absorption minima in the vicinity of the reference wavelengths for the lines of interest. The detected peaks provide initial estimates of the central wavelength, line depth, and full width at half maximum (FWHM). Around each candidate line centre, the algorithm defines a small local fitting window and performs a non-linear least-squares fit of a Voigt function to the observed spectrum. The continuum/baseline level is therefore treated locally for each fitted line rather than as a single value across the full 327--329\,nm spectral window, which is essential in this dense near-UV region where neighbouring lines and local profile asymmetries are common. The Voigt profile combines Gaussian and Lorentzian functions, allowing us to account, in an effective way, for thermal, instrumental, and pressure broadening (for saturated lines). 

A key feature of the code is its adaptive fitting strategy: the initial line depth and FWHM estimates are used to set line-specific parameter bounds and, where needed, to adjust the fitting window and weighting. This allows the same routine to robustly fit many lines simultaneously, despite their differing depths, widths, and profile shapes, and helps to avoid unphysical broadening or spurious shifts for weak or blended lines.

The fitting is performed iteratively. After an initial fit, simultaneously to all selected lines, the resulting line centres and widths are used as updated initial guesses for a second fitting round. This process is repeated until the change in the total residual (summed over all fitted lines) and in the individual line centres falls below a chosen convergence threshold, or until a maximum number of iterations is reached. We also impose moderate bounds on the allowed shifts and widths to prevent the fits from drifting to unphysical solutions in noisy profiles.

For each line, the final fit provides (1) the best-fit central wavelength, $\lambda_{\mathrm{core}}$, (2) the line-core intensity, $I_{\mathrm{core}}$, taken as the fitted line-core intensity evaluated at $\lambda_{\mathrm{core}}$, and (3) an associated LOS velocity, $v_{\mathrm{LOS}} = c\,\frac{\lambda_{\mathrm{core}} - \lambda_{0}}{\lambda_{0}}$, where $\lambda_{0}$ is the reference wavelength and $c$ is the speed of light.

\begin{figure*}[!ht]
\centering
\includegraphics[width=1.0\textwidth]{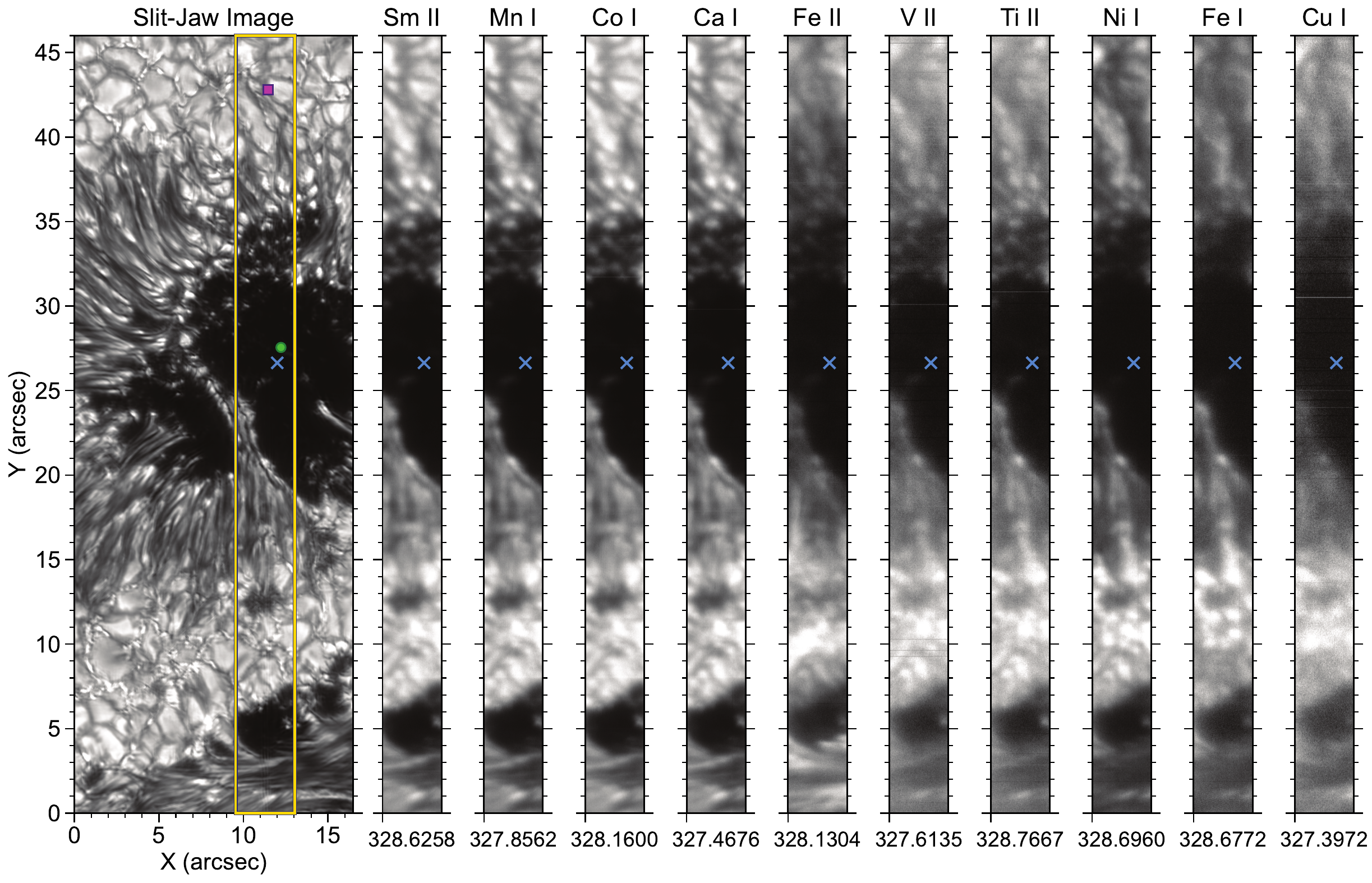}
\caption{Context for the umbral analysis region. 
Left: SUSI slit-jaw image showing the sunspot and its surroundings at a (near-)continuum position. The yellow solid line marks the small raster scan area and the blue cross indicates the location in the central umbra used for the time-series analysis. The filled magenta square marks the additional weak-field reference pixel outside the umbra, while the filled green circle marks the secondary umbral-core pixel analysed in Appendix~\ref{sec:appendix}.
Right: example line-core intensity maps for a subset of the selected near-UV lines, ordered from the deep photosphere to the low chromosphere (left to right), illustrating how the same spatial region is simultaneously sampled at different effective heights. The spectral lines' wavelengths (in nm) are indicated below each raster image.
}
\label{fig:umbra_context}
\end{figure*}

The procedure yields sub-pixel accurate line centres and robust line-core intensities and LOS velocities for all 44 lines at each time step and spatial position. In this Letter we focus on a small $3\times 3$ pixel region centred on the umbral core, where the magnetic field is expected to be relatively vertical and height comparisons are therefore most straightforward. For this region, we first average the Stokes~$I$ spectra over the $3\times 3$ pixels to improve the signal-to-noise ratio, and then apply the line-fitting procedure to the binned profiles. The analysis location is close to the umbral centre, where horizontal gradients are weak, so this spatial binning does not significantly smear the underlying structure. This yields time series of $I_{\mathrm{core}}(t)$ and $v_{\mathrm{LOS}}(t)$ for each spectral line.
An example of the umbral analysis location and its surroundings is shown in Figure~\ref{fig:umbra_context}, which displays the SUSI slit-jaw image with the analysis region marked, together with representative intensity maps for a subset of the selected lines. Please note that the difference in sharpness between the slit-jaw image and the spectral scans is because the former has been phase-diversity reconstructed while the latter have not. 

\subsection{Wavelet-based Power Spectra}
\label{subsec:wavelet}

To characterise the wave content of the line-core time series, we use the wavelet analysis framework implemented in the WaLSAtools\footnote{\url{https://github.com/WaLSAteam/WaLSAtools}} library \citep{2025NRvMP...5...21J, walsatools..2025...17569951}. For each spectral line, and for both $I_{\mathrm{core}}(t)$ and $v_{\mathrm{LOS}}(t)$, we compute the Morlet wavelet transform and the associated wavelet power spectrum as a function of time and frequency. Following standard practice, we remove a linear trend from each time series before analysis and apodise the signal using a Tukey window with a 10\% taper. In addition, we suppress very low-frequency components ($\nu \lesssim 2$\,mHz), which mainly reflect slow background evolution rather than the wave dynamics of interest here. This is done via wavelet-based filtering: we reconstruct a filtered version of each time series using the inverse wavelet transform while retaining only coefficients with $\nu > 2$\,mHz (\texttt{recon} routine in WaLSAtools), and perform the subsequent power analysis on the filtered signals. Representative examples of these time series are shown in Appendix~\ref{sec:appendix_timeseries}.
 
Significance levels are estimated from one thousand randomised surrogate signals, and we define the cone of influence (COI) to identify regions of each wavelet spectrum affected by edge effects. To obtain a compact measure of the frequency distribution of power, we compute the refined global wavelet spectrum (RGWS) for each time series. The RGWS is defined as the time-integrated wavelet power as a function of frequency, excluding regions affected by the COI and retaining only those frequencies where the local power exceeds the 95\% confidence level \citep{2025NRvMP...5...21J}. This yields a one-dimensional spectrum that captures the dominant oscillatory power while reducing sensitivity to edge artefacts and to fluctuations that do not pass the adopted confidence threshold. We verified that lowering the threshold to 90\% does not change the main qualitative results, but mainly introduces weaker and noisier additional features. The resulting spectra are generally smooth, and show excellent consistency with classical FFT power spectra, while providing better localisation of transient and quasi-periodic behaviour.

\begin{figure*}[!ht]
\centering
\includegraphics[width=1.0\textwidth]{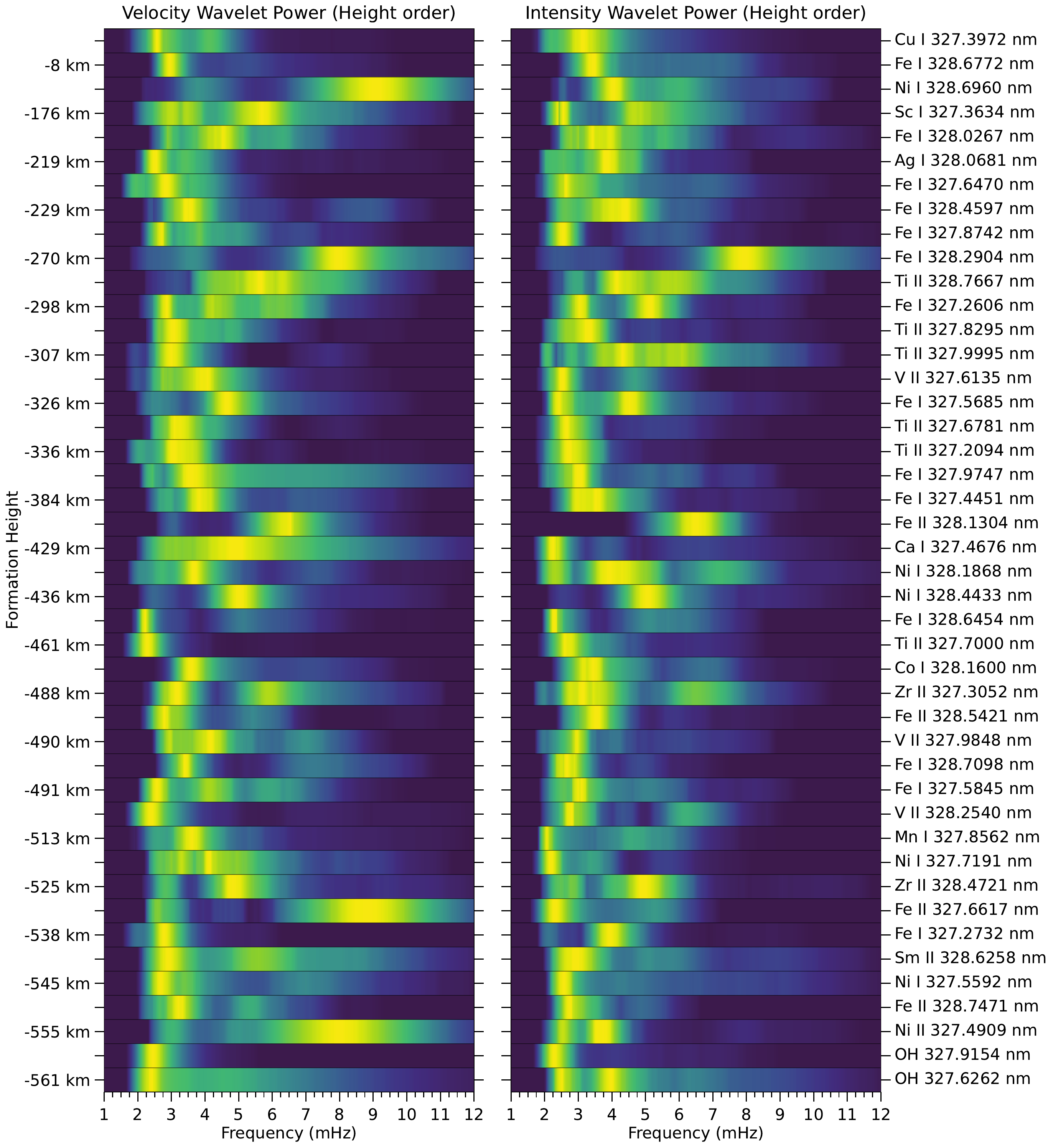}
\caption{Stacked refined global wavelet spectra (RGWS) for the 44 selected near-UV lines in the umbral core, ordered by their effective line-core formation heights.
Left: normalised RGWS for line-of-sight velocity; Right: normalised RGWS for line-core intensity. 
In each panel, the horizontal axis is frequency, the vertical axis is the effective line-core formation height from the sunspot model (with the corresponding spectral lines labelled on the right), and the colour scale shows the RGWS power normalised by the maximum of each individual spectrum.} 
\label{fig:rgws-power-spectra}
\end{figure*}

For the umbral core region, this procedure yields, for each of the 44 lines, two RGWS spectra: one for intensity and one for LOS velocity. To enable a meaningful comparison across diagnostics with different absolute power levels, each RGWS is normalised by its own maximum value. We then stack the spectra into two 2D diagrams (one for LOS velocity and one for intensity), with frequency on the horizontal axis and the vertical axis ordered by the estimated line-core formation height derived from the sunspot model atmosphere. The resulting height-ordered RGWS maps are shown in Figure~\ref{fig:rgws-power-spectra}; the corresponding spectral lines are labelled along the right-hand side, and the colour scale represents the normalised power. This model-based ordering is used as a geometrical proxy to explore possible trends with height, while recognising that the absolute values, and in some cases the relative ranking of individual lines, remain model dependent or uncertain. 

We note that this normalisation removes information on absolute power variations between lines, which may carry physical meaning (e.g., amplitude scaling and energy-flux estimates). A quantitative interpretation of absolute power, however, requires better-constrained formation heights and atmospheric parameters, and is therefore deferred to future work using full-Stokes inversions.

Figure~\ref{fig:rgws-power-spectra} summarises how the distribution of oscillatory power varies across the ensemble of near-UV diagnostics and highlights differences between intensity- and velocity-based signals. Rather than a single preferred band, the individual spectra show several significant power enhancements across the frequency range of 2--12\,mHz: some lines display a single dominant peak, others show enhanced power at two or more distinct frequencies, and the locations of the main peaks change from line to line. The apparent ``jumps'' in dominant frequency between neighbouring lines therefore point to a genuinely diverse set of frequency responses, rather than a simple monotonic trend with the model-based height ordering. We emphasise that the occurrence of statistically significant power up to $\sim$8--10\,mHz in some diagnostics is an empirical property of these RGWS spectra; interpreting such components in terms of cut-off behaviour, resonance, or specific mode content is deferred to follow-up work with improved height calibration and full-Stokes constraints.

To obtain a complementary view that does not rely on the model height scale, we also applied an unsupervised machine-learning clustering analysis to the RGWS spectra. Specifically, we used hierarchical agglomerative clustering (cosine distance, average linkage; \citealt{2012WDM..2....86}) primarily as a shape-based reordering method for the normalised power-versus-frequency profiles in the 1--12\,mHz range, separately for LOS velocity and for intensity. In practice, the clustering groups spectra according to the similarity of their overall power-distribution shapes, after which the resulting groups are arranged from lower- to higher-characteristic-frequency content for display. This provides a compact, height-independent organisation of the multi-peaked spectra without imposing any prior assumptions about formation height. We verified that this reordered organisation is robust to the normalisation choice: repeating the clustering on unnormalised RGWS spectra yields essentially the same qualitative spectral grouping and line ordering.

\begin{figure*}
  \centering
  \includegraphics[width=\linewidth]{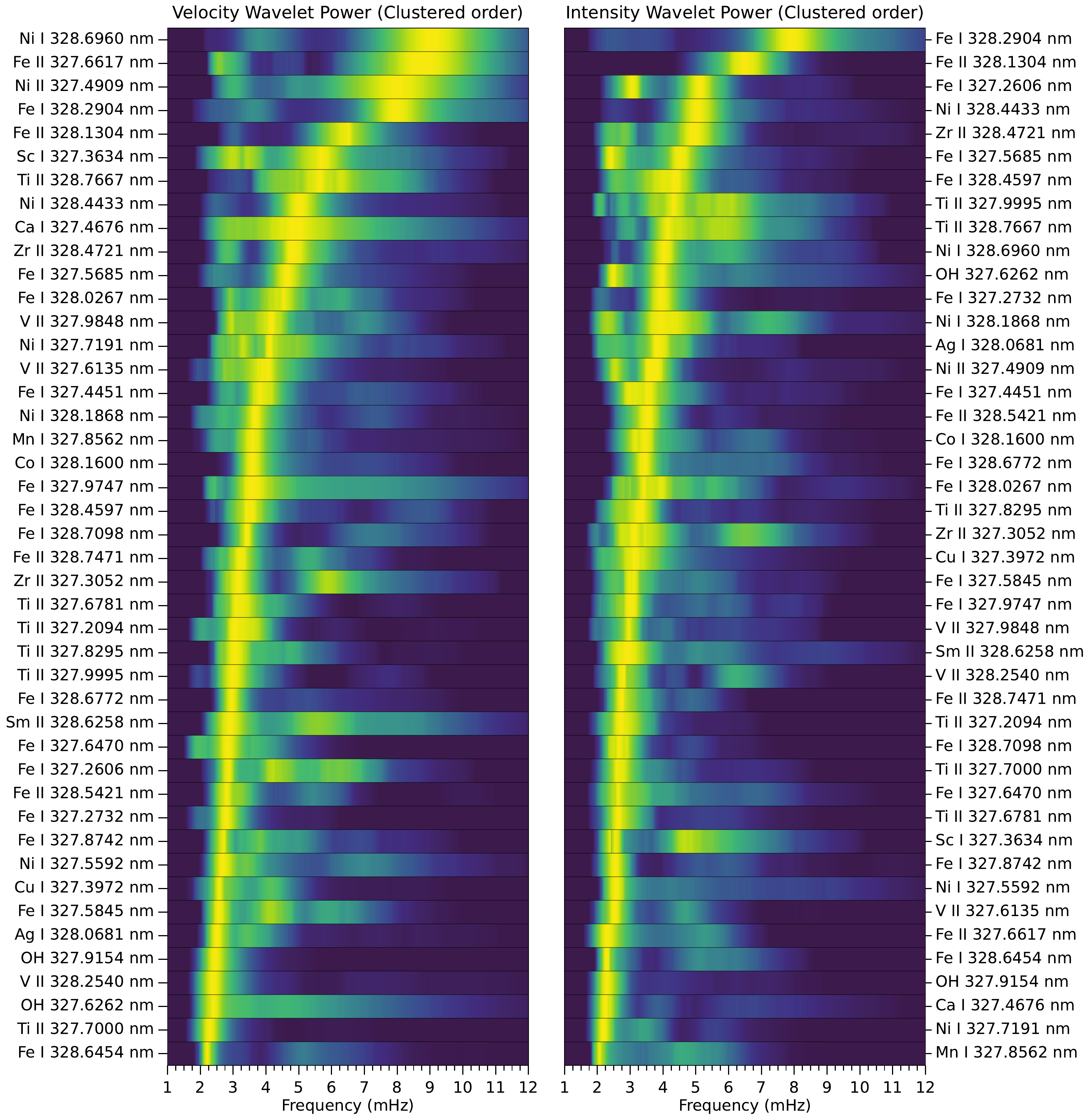}
  \caption{Same refined global wavelet spectra as in Figure~\ref{fig:rgws-power-spectra}, but reordered using hierarchical clustering of the normalised RGWS profiles. 
  Left: line-of-sight velocity; Right: line-core intensity. 
  In each panel, rows correspond to individual spectral lines, reordered so that spectra with similar overall power-distribution shapes are displayed together, with the resulting groups arranged from lower- to higher-characteristic-frequency content for visual comparison. This height-independent reordering highlights differences between intensity- and velocity-based wave signatures.}
  \label{fig:rgws_clustered}
\end{figure*}

The clustering-based reordering is visualised in Figure~\ref{fig:rgws_clustered}, where the same RGWS spectra are now displayed according to the similarity of their overall power-distribution shapes. For visual clarity, the resulting groups are arranged from lower- to higher-characteristic-frequency content, so that the reordered stacks provide a convenient progression across the ensemble. Because many diagnostics display multiple significant peaks, this approach is not equivalent to sorting by the dominant peak alone: spectra are first grouped by their overall multi-peaked shape (e.g., the presence, absence, and relative strength of secondary bands), and the final display order is then chosen to provide an intuitive low- to high-frequency progression. This height-independent organisation highlights that similar multi-peaked spectral shapes recur across the ensemble, and that intensity and velocity organise into somewhat different families.

The stacked RGWS spectra show that the umbral oscillations are genuinely multi-frequency and that, across the ensemble of near-UV lines, the dominant power peak spans a broad range of frequencies (roughly $\sim$2--10\,mHz). This spread is most readily seen in the cluster-ordered view, which provides a height-independent organisation of the same spectra by similarity of their power-distribution shapes. At this stage, however, we cannot yet cleanly separate the roles of uncertain formation heights and line-specific radiative sensitivities in shaping the line-to-line differences.

\section{Conclusions} 
\label{sec:conclusions}

In this Letter we have presented the first systematic multi-line study of sunspot wave signatures in the near-UV, using spectral lines observed by \textsc{Sunrise iii} /SUSI in the 327--329\,nm window. Exploiting a dense set of diagnostics, we extracted line-core intensity and LOS velocity time series for 44 carefully selected, relatively clean lines (excluding profiles with peculiar/highly dynamic umbral behaviour or strongly skewed cores suggestive of potential blends) and analysed their oscillatory properties in a small umbral core region. Sub-pixel line-core measurements obtained from an adaptive Voigt-fitting routine were combined with Morlet wavelet analysis to construct RGWS that summarise the frequency content of the oscillations across the ensemble of near-UV lines. By construction, the RGWS retain only power outside the cone of influence and above the 95\% confidence level, so power at multiple frequencies -- including features substantially weaker than the absolute maximum -- represents statistically significant components of the signal. Previous studies have shown that such weaker components, when isolated through appropriate frequency filtering or advanced time–frequency analysis, can correspond to physically meaningful MHD wave modes rather than noise \citep[e.g.][]{2017ApJ...842...59J, 2022NatCo..13..479S, 2024A&A...688A...2J}. 

The RGWS diagrams reveal that the umbral oscillations sampled by these diagnostics are intrinsically multi-frequency. Many lines display several significant power enhancements between $\sim$2 and 12\,mHz, and the locations of their strongest peaks vary from line to line. When the spectra are ordered using model-based estimates of line-core formation height, no simple monotonic trend emerges in which a single frequency band gradually replaces another with increasing height. Instead, low-, mid-, and high-frequency power enhancements are mixed throughout the set of diagnostics. Consistently, we find no clear ordering when grouping the spectra by basic line properties such as line depth or simple measures of contribution-function width, indicating that these simple proxies do not isolate the frequency progression. This behaviour could indicate that (\textit{i}) the assumed height scale does not fully capture the true ordering of the line-formation layers, or that (\textit{ii}) line-dependent diagnostic responses (e.g. differing sensitivities to temperature, velocity, and density, as well as varying degrees of scattering) modulate how a common underlying wave field is expressed in each observable. A further contributor may also be uncertainties in atomic line parameters (e.g., oscillator strengths and damping constants) in the near-UV, which can propagate into radiative-transfer-based height estimates and response functions. Because contribution functions depend on both opacity and the source function, scattering-sensitive lines (i.e. affected by non-LTE radiative transfer) can introduce line-dependent biases in the inferred formation heights when these are estimated under LTE assumptions, in addition to producing different diagnostic responses to the same underlying perturbations \citep[e.g.][]{2008A&A...480..515C, 2012ApJ...749..136L}. Additional uncertainty may arise from blends that become important in the umbra but are not obvious from quiet-Sun atlas spectra; while we excluded lines with visually suspicious umbral profiles, subtle blends may still affect some diagnostics and their effective formation heights. At this stage, the analysis does not allow these effects to be cleanly disentangled, but it does show that a simple ``one dominant frequency per height'' picture is inadequate; the more robust conclusion is that different near-UV diagnostics respond in systematically different ways to the same sunspot oscillations. At the same time, the presence of multiple significant peaks does not by itself establish harmonic relationships. The observed peak spacings are not consistent with a simple harmonic sequence, and demonstrating harmonics or resonant-cavity modes would in any case require additional constraints, such as phase relations, spatial mode structure, and multi-height coherence analyses \citep[see, e.g.,][]{2020NatAs...4..220J, 2021NatAs...5....5J, 2020ApJ...900L..29F, 2024MNRAS.529..967S, 2025ApJ...986..180S}, which are beyond the scope of the present Letter.

Multi-frequency umbral spectra -- typically a dominant $\sim3-6$\,mHz band with secondary peaks that can extend to $\sim8-12$\,mHz -- have been reported in a range of single-line observations \citep{2014RAA....14.1458R, 2019A&A...621A..43F, 2023A&A...674A.109C, 2025A&A...697A.156B}. Such multi-peak spectra have, in particular, been interpreted as signatures of multiple coexisting sunspot eigenmodes in recent analyses \citep[e.g.][]{2022NatCo..13..479S}. Our results are consistent with this overall frequency landscape, but the key advance here is the multi-line diagnostic mapping: dozens of near-UV lines provide simultaneous observables with different radiative sensitivities (to velocity, temperature, and opacity/source-function variations) and different effective formation depths, offering much richer constraints on how the multi-peak spectra manifest themselves across the photosphere and the low chromosphere than is possible with one or two lines. A complementary multi-line result has also recently been reported from the ground. \citet{2026FrASS...Grant_inprep} analyse 21 spectral lines observed with the FRANCIS fibre-fed integral-field unit \citep{2023SoPh..298..146J} at the Dunn Solar Telescope in the 580.7--597.3\,nm range. They likewise find that the dominant oscillation peaks in the line-core LOS-velocity time series vary across the diagnostics in the umbral core, with significant power spanning 3--12\,mHz, despite the different instrument, spectral window, and velocity-estimation methodology used compared to the present work.

We also find that intensity- and velocity-based diagnostics can display markedly different spectral shapes for the same nominal line. Some diagnostics show broadly similar dominant peaks in both $I_{\mathrm{core}}$ and $v_{\mathrm{LOS}}$, while others exhibit pronounced power at certain frequencies in one quantity but not the other. Rather than a single systematic pattern, the near-UV lines sample a diverse mix of responses, again pointing to a combination of physical wave properties and line-formation effects. In general, $v_{\mathrm{LOS}}$ is most directly sensitive to Doppler shifts, whereas $I_{\mathrm{core}}$ reflects a combination of opacity and source-function variations and can therefore be more sensitive to thermodynamic perturbations (while still retaining some sensitivity to velocity through line-shape changes). Scattering-dominated lines can further weight the signal non-locally, potentially enhancing or suppressing specific frequency components compared to more LTE-like lines. In addition, in many lines, Doppler-based velocities often have substantial sensitivity contributions from the line flanks, whereas line-core intensity is most sensitive close to the line-core formation region; thus the two observables can sample different height ranges even within the same line. These contrasts underscore that intensity and velocity diagnostics need not trace the same wave components, even within the same line, because they sample different combinations of perturbations through line formation. A full exploitation of these diagnostic differences will require more detailed radiative-transfer modelling than attempted here.

To further probe the diversity of frequency responses, we complemented the height-ordered stacks with an unsupervised clustering analysis of the RGWS spectra. The cluster-ordered diagrams group lines with similar multi-peaked power distributions and, when arranged by characteristic frequency, reveal families of diagnostics in which the dominant power shifts progressively from $\sim$2 to $\sim$10\,mHz across the ensemble, for both intensity and velocity (though not necessarily in the same lines). These frequency-based groupings do not map one-to-one onto the model height ordering, which reinforces the need for better-constrained formation heights and more detailed radiative-transfer calculations, including a careful treatment of blends that become important in cool umbral conditions. At the same time, the clustering results strengthen the conclusion that the multi-frequency structure seen in the near-UV diagnostics is an intrinsic property of the way these lines sample the sunspot atmosphere, rather than an artefact of a particular choice of height scale. As an additional validation, repeating the same analysis for several random pixels outside the umbra does not reproduce the frequency progression seen in the umbral core, whereas sampling a few nearby umbral-core locations yields qualitatively similar behaviour. This supports the interpretation that the observed frequency-structured organisation is intrinsic to the sunspot umbra rather than a systematic instrumental or calibration artefact. Examples of the corresponding RGWS stacks for an additional umbral-core pixel and for a representative weak-field reference location outside the umbra are shown in Appendix~\ref{sec:appendix}.

The present work is intentionally conservative, focusing on a single umbral location and on robust RGWS diagnostics. This spatial location was chosen because the magnetic field there is expected to be strong and relatively vertical compared with surrounding regions, so that LOS velocities and intensities provide the cleanest practical proxy for vertically stacked diagnostics and minimise geometric complications. We note, however, even in the umbral core, small residual inclinations and height-dependent expansion can cause different lines to sample oscillations along slightly different field lines, potentially contributing to line-to-line differences. For the relatively small inclinations expected in the umbral core, and over the limited height range sampled here, such effects are not expected to qualitatively alter the main wave-signature organisation discussed in this Letter, although a quantitative inclination- and cut-off-based interpretation is deferred to future work once full-Stokes inversions and more rigorous height-sensitivity constraints become available. This Letter provides a first survey of how oscillatory signatures appear across a large set of near-UV lines in a sunspot, laying the groundwork for more detailed mode identification and energy-budget studies. In future work, we will extend this analysis systematically to multiple positions within and around the sunspot, and complement the RGWS diagnostics with phase and coherence measurements between neighbouring lines to probe propagation, reflection, and possible resonance signatures. A key next step is to better constrain the height sensitivity of the SUSI near-UV diagnostics using detailed radiative-transfer response functions for velocity and temperature, complemented by multi-line (full-Stokes) inversions \citep[e.g.][]{2019A&A...622A..36R, 2026A&A...705A.220H} once polarimetric products become available. Response functions provide a direct line-by-line measure of where each diagnostic is most sensitive (and over what height range), while inversions can recover stratified atmospheric parameters, including magnetic-field strength and inclination variations needed for a fuller interpretation of the observed frequency structure in terms of specific MHD modes and propagation paths. Complementary 3D radiative-MHD simulations can then be used for forward modelling to test these interpretations and to link the observed multi-frequency spectra to wave modes, resonance structures, and energy-transport pathways.

More advanced data-driven methods (e.g. proper orthogonal decomposition, empirical mode decomposition with Hilbert-based diagnostics, and synchrosqueezing transforms) will also be explored to isolate recurrent patterns and candidate modes more quantitatively. Ultimately, combining the near-UV diagnostics with SCIP \citep{2026arXiv260317929K} and TuMag \citep{2025SoPh..300..148D} observations will extend the coverage across the photosphere–chromosphere system and enable a more complete multi-height interpretation.

\begin{acknowledgments}
SJ and DBJ acknowledge support from the UK Science and Technology Facilities Council (STFC) through consolidated grants ST/T00021X/1 and ST/X000923/1. SJ also received support from the Rosseland Centre for Solar Physics (RoCS), University of Oslo, Norway. DBJ further acknowledges funding from the Leverhulme Trust (Research Project Grant RPG-2019-371) and from the UK Space Agency via the National Space Technology Programme (grant SSc-009).
RJM is supported by the UKRI Future Leaders Fellowship (RiPSAW—MR/T019891/1 and MR/Z000289/1). TF acknowledges grants PID2021-127487NB-I00, PID2024-156538NB-I00, CNS2023-145233, and RYC2020-030307-I funded by MCIN/AEI/10.13039/501100011033. We wish to acknowledge scientific discussions with the Waves in the Lower Solar Atmosphere (WaLSA; \href{https://WaLSA.team}{www.WaLSA.team}) team, which has been supported by the Research Council of Norway (project no. 262622), The Royal Society (award no. Hooke18b/SCTM; \citealt{2021RSPTA.37900169J}), and the International Space Science Institute (ISSI Team 502).
\textsc{Sunrise iii}  is supported by funding from the Max-Planck-F\"orderstiftung (Max Planck Foundation), NASA under Grants \#80NSSC18K0934 and \#80NSSC24M0024 (``Heliophysics Low Cost Access to Space'' program), and the ISAS/JAXA Small Mission-of-Opportunity program and JSPS KAKENHI Grant Numbers JP18H05234 and JP23K25916. This research has received financial support from the European Union's Horizon 2020 research and innovation programme under grant agreement No.~824135 (SOLARNET) and No.~101097844 (WINSUN) from the European Research Council (ERC). It has also been funded by the Deutsches Zentrum f\"ur Luft- und Raumfahrt e.V.\ (DLR, grant no.~50~OO~1608). The Spanish contributions have been funded by the Spanish MCIN/AEI under projects RTI2018-096886-B-C5 and PID2021-125325OB-C5, and from ``Center of Excellence Severo Ochoa'' awards to IAA-CSIC (SEV-2017-0709, CEX2021-001131-S), all co-funded by European REDEF funds, ``A way of making Europe''.
\end{acknowledgments}

\appendix

\section{Representative LOS-velocity and line-core intensity time series}
\label{sec:appendix_timeseries}

\begin{figure*}[!ht]
  \centering
  \includegraphics[width=\linewidth]{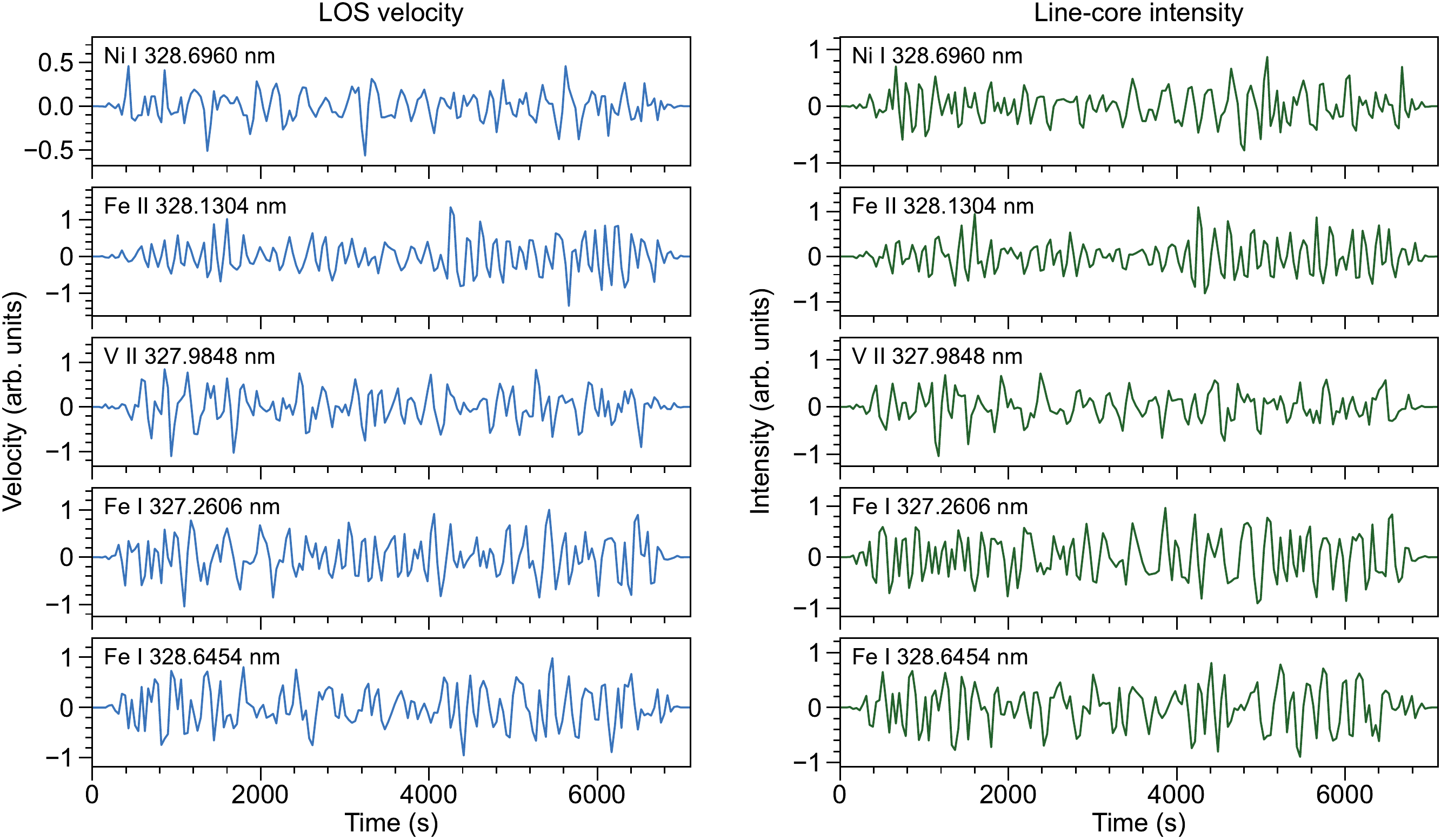}
  \caption{Representative line-core intensity and LOS-velocity time series for a small subset of the analysed near-UV lines in the primary umbral-core location.}
  \label{fig:timeseries_examples}
\end{figure*}

To illustrate the temporal behaviour underlying the wavelet-based analysis, Figure~\ref{fig:timeseries_examples} shows representative LOS-velocity and line-core intensity time series for five selected lines in the primary umbral-core location. The signals shown are the same detrended, apodised, and low-frequency-filtered time series used in the wavelet analysis described in Section~\ref{subsec:wavelet}. Because this preprocessing is designed to isolate the oscillatory content of interest, it can also affect the absolute amplitude scale to some extent; the plotted amplitudes are therefore shown in arbitrary units. This does not affect either the analysis or its interpretation in the present work, since our focus is on the frequency structure and relative organisation of the oscillatory signals rather than on absolute amplitudes or absolute power comparisons between lines. These examples are included to illustrate the temporal behaviour, intermittency, and relative visibility of different oscillatory components.

\section{Additional RGWS examples: a weak-field reference location outside the umbra} and a secondary umbral location
\label{sec:appendix}

To illustrate that the frequency-structured organisation reported in the main text -- in particular the frequency progression revealed by the clustering-based reordering in Figure~\ref{fig:rgws_clustered} -- is characteristic of the umbra, we repeated the same RGWS and hierarchical-clustering analysis for (i) a weak-field reference pixel outside the umbra within the raster field of view and (ii) an additional umbral-core pixel close to the primary analysis location. Both supplementary locations are indicated in the slit-jaw context image (Figure~\ref{fig:umbra_context}) using different marker colours.

\begin{figure*}[t!]
  \centering
  \includegraphics[width=\linewidth]{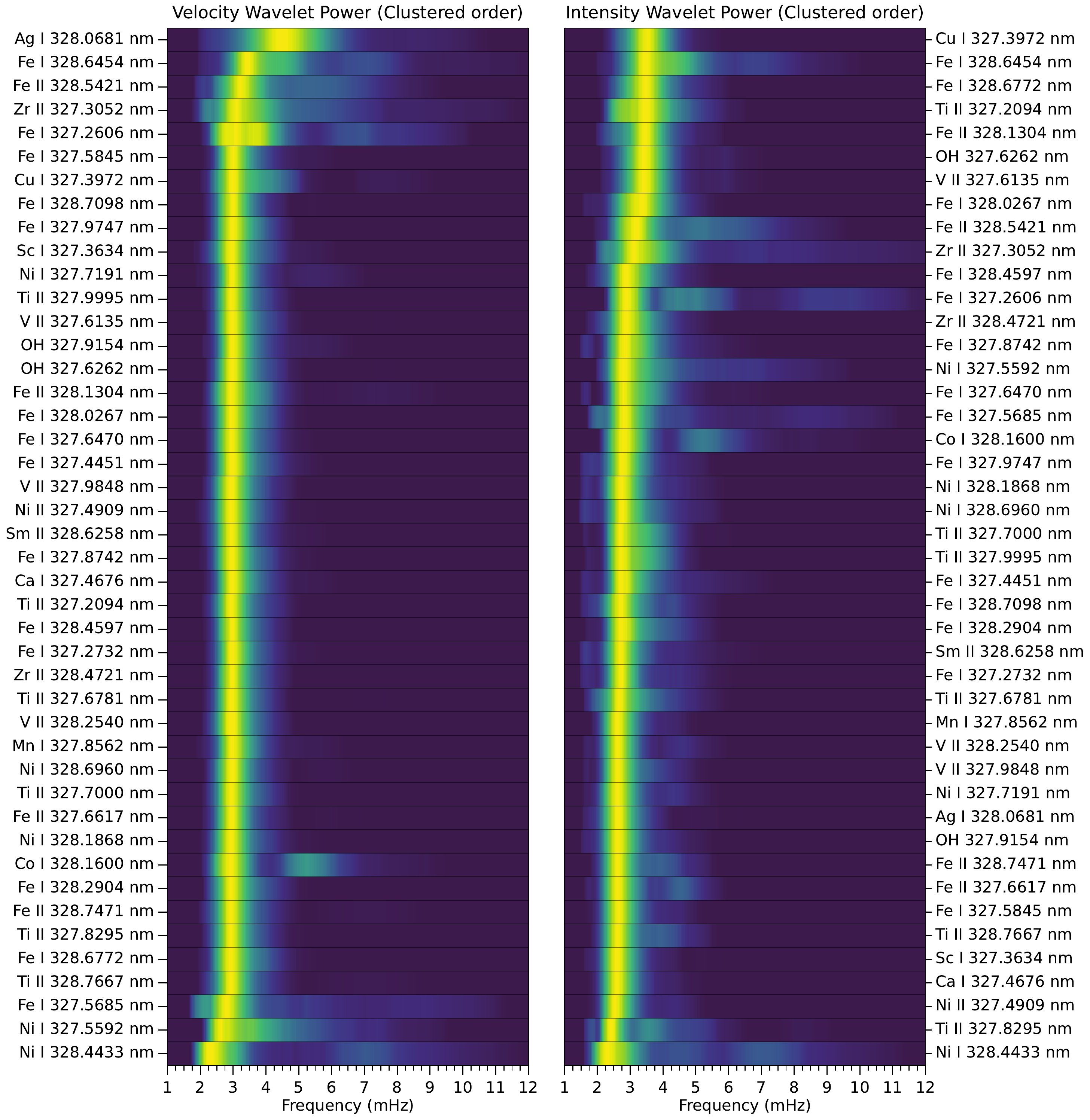}
  \caption{Cluster-ordered, normalised refined global wavelet spectra (RGWS) for a representative weak-field reference pixel outside the umbra marked by a filled magenta square in Figure~\ref{fig:umbra_context}, shown in the same format as Figure~\ref{fig:rgws_clustered}. Left: line-of-sight velocity; Right: line-core intensity. Rows correspond to individual spectral lines, reordered using hierarchical clustering of the normalised RGWS profiles and arranged by dominant-frequency content.}
  \label{fig:rgws_clustered_qs}
\end{figure*}

In the weak-field reference example outside the umbra (Figure~\ref{fig:rgws_clustered_qs}), the clustering-based reordering does not reveal a comparable progression of dominant frequencies across the line ensemble. Instead, most lines concentrate their dominant power near $\sim$3.5\,mHz (within $\pm0.5$\,mHz), with only minor deviations in a small number of diagnostics. This contrast supports the interpretation that the structured frequency organisation seen in the umbral core is not a generic property of the analysis procedure or the instrument, but is linked to the sunspot umbra.

\begin{figure*}[!ht]
  \centering
  \includegraphics[width=\linewidth]{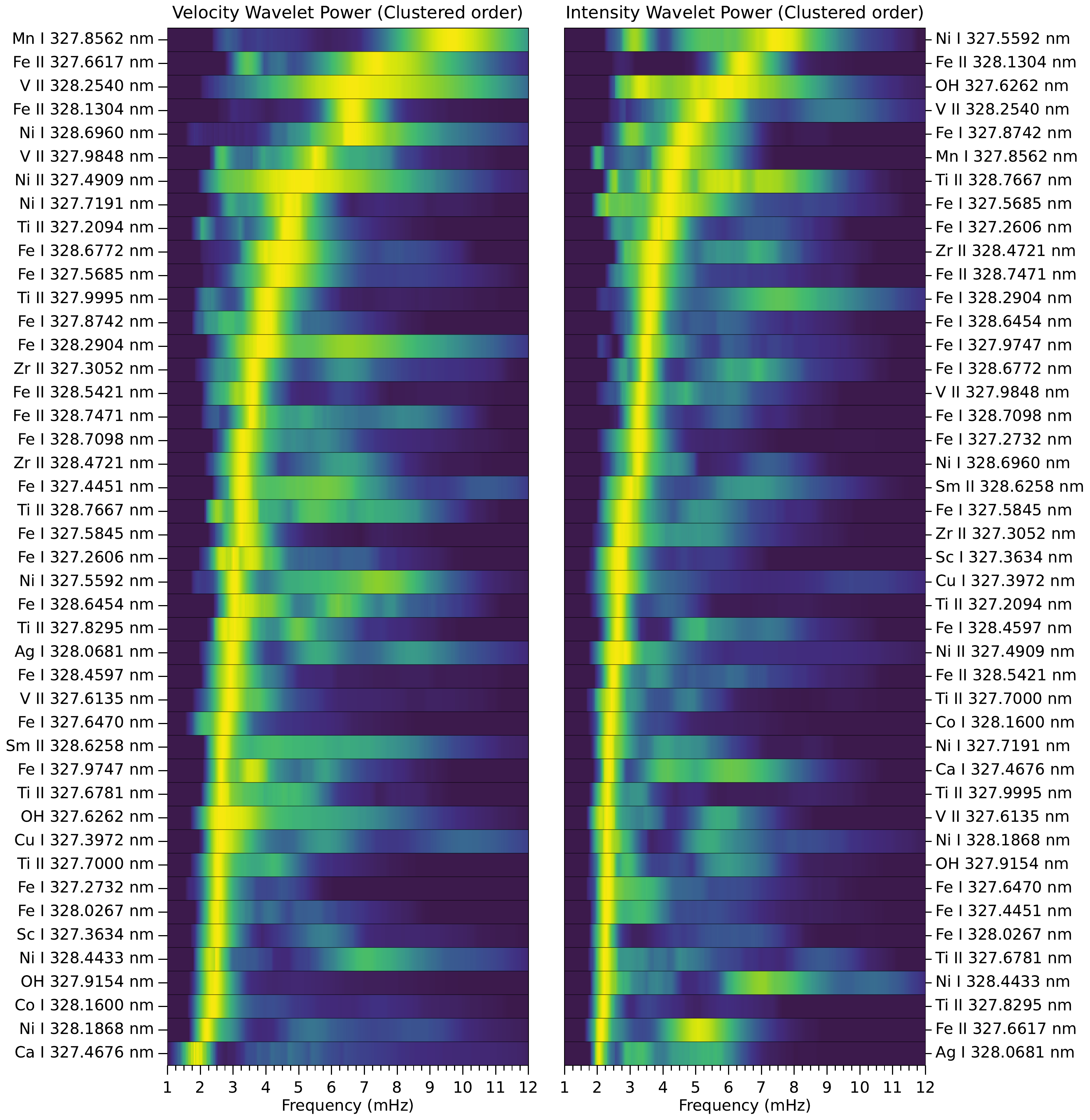}
  \caption{Same as Figure~\ref{fig:rgws_clustered_qs}, but for an additional umbral-core pixel marked by a filled green circle in Figure~\ref{fig:umbra_context}. While the detailed line ordering can vary between neighbouring umbral pixels, the qualitative progression of dominant frequencies across the ensemble is recovered.}
  \label{fig:rgws_clustered_umb2}
\end{figure*}

For the secondary umbral-core pixel (Figure~\ref{fig:rgws_clustered_umb2}), we recover qualitatively similar behaviour to the primary umbral location: the cluster-ordered stacks again show a progression of dominant frequencies spanning roughly $\sim$2--10\,mHz, for both LOS velocity and intensity (though not necessarily in the same diagnostics). The detailed line ordering and, for some lines, the dominant peak frequency differ from the primary umbral pixel. Such differences are expected: even within the umbral core, small lateral variations in atmospheric structure and magnetic configuration (e.g. field strength/inclination, thermodynamic stratification, and flows), together with height-dependent dynamics, can shift line-formation weighting and the relative visibility of coexisting oscillatory components, such that the strongest peak in a multi-peak spectrum may change between neighbouring pixels. Consistent with this, the 3D MURaM sunspot model shows substantial pixel-to-pixel lateral variations in effective formation height within the umbra across the 327--329\,nm range, with a typical span of several hundred kilometres (median $\sim$615\,km) and extremes exceeding 1\,Mm. Therefore, an identical cluster-based line ordering is not expected from one umbral pixel to another, even when the overall multi-frequency character and the qualitative frequency progression persist.

\bibliography{article}{}
\bibliographystyle{aasjournalv7}

\end{document}